\documentclass[12pt,twoside]{article}

\usepackage{epsfig}
\usepackage{amsmath}

\newcommand{\poin}{Poincar\'e}
\newcommand{\pderiv}[2]{\frac{\partial #1}{\partial #2}}
\newcommand{\deriv}[2]{\frac{d#1}{d#2}}
\newcommand{\eol}{\notag\\}

\newcommand{\GeVA}{\text{GeV$\cdot$A}}

\begin{document}

\begin{titlepage}
  \begin{flushright}
    hep-ph/9907215
  \end{flushright}
  \vspace{3cm}
  \begin{centering}
  
   {\Large \textbf{A Poincar\'e-Covariant Parton Cascade Model for
       Ultrarelativistic Heavy-Ion Reactions\footnote{ Talk given by V.
         B\"orchers in \emph{Quark Matter '99}, May 13, 1999, Torino,
         Italy. For the transparencies, see
         http://www.qm99.to.infn.it/program/qmprogram.html}}\\ }
   
   \vspace{0.5cm}
   {\large V.\ B{\"o}rchers, C.C.\ Noack\\ }
   \vspace{0.3cm}
   \emph{Institut f{\"u}r Theoretische Physik
     Universit{\"a}t Bremen,\\ D--28334 Bremen, Germany\\
     e-mail: \texttt{boercher,noack@physik.uni-bremen.de\\ }}
   
   \vspace{0.3cm}
   {\large S.\ Gieseke\\ }
   \vspace{0.3cm}
   \emph{Now at II.\ Institut f\"ur Theoretische Physik,
     Universit\"at Hamburg, Germany\\
     e-mail: \texttt{Stefan.Gieseke@desy.de\\ }}
   
   \vspace{0.3cm}
   {\large G.\ Martens\\ }
   \vspace{0.3cm}
   \emph{Now at Institut f\"ur Theoretische Physik,
     Universit\"at Giessen, Germany\\
     e-mail: \texttt{Gunnar.Martens@theo.physik.uni-giessen.de\\ }}
   
   \vspace{0.3cm}
   {\large J.\ Meyer\\ }
   \vspace{0.3cm}
   \emph{Now at Institut f\"ur Theoretische Physik,
     Universit\"at Heidelberg, Germany\\
     e-mail: \texttt{jochen@tphys.uni-heidelberg.de\\ }}
   \vspace{1cm}
 \end{centering}

\begin{abstract}
We present a new cascade-type microscopic simulation of Nucleus-Nucleus
Collisions at RHIC energies. The basic elements are partons (quarks and
gluons) moving in $8N$-dimensional phase space according to
\poin-covariant dynamics. The parton-parton scattering cross sections
used in the model are computed within perturbative QCD in the tree-level
approximation.  The $Q^2$ dependence of the structure functions is
included by an implementation of the DGLAP mechanism suitable for a
cascade, so that the number of partons is not static, but varies in
space and time as the collision of two nuclei evolves.  The resulting
parton distributions are presented and compared to distributions
obtained with the VNI model.
\end{abstract}
\end{titlepage}


\section{Introduction}
\label{sec:intro}
For a long time nuclear reactions have been described by
phenomenological hadronic models quite successfully. At nuclear
collision energies of the order of $10 \GeVA$ it is already clear that
partonic degrees of freedom will not be negligible; thus pure hadron
cascades do not suffice in the SPS regime.

Parton cascade models on the other hand provide a microscopic
description of high energy collisions in terms of partons.  The
QCD-model of \emph{hard scatterings}, the theoretical basis of parton
cascades, limits them to reactions with high transferred momentum. The
large number of involved particles (nucleons) in heavy ion collisions
lets one assume that not all single nucleon-nucleon collisions will be
hard. However, for sufficiently high energies of the incident nuclei one
can expect that hard collisions are dominant.

In the framework of a parton cascade non-perturbative effects can be
either completely neglected or treated phenomenologically. For the model
to be presented here we have chosen the first alternative, although this
makes the model not applicable in the SPS energy regime.

The first parton cascade model for heavy ion collisions, HIJING
\cite{Wang:1991ht}, was formulated in momentum space only, and thus did
not include a space-time description of the cascade evolution. A full
phase-space description was first provided by Klaus Geiger's VNI
\cite{Geiger:1992nj,Geiger:1995he}, but this model is not manifestly
Poincar\'e-invariant.

A description of the space-time evolution of the system is imperative if
one aims to include effects due to the high density produced in the
$A-A$ collision.

\section{Parton cascades and \poin-covariance}

The obvious way to enhance a parton cascade with a space-time
description would be to choose an observer Lorentz frame and then demand
that two particles $i$ and $j$ interact instantaneously as seen in this
frame. It is clear that such a procedure makes the ordering of the
interactions frame-dependent. This has been recognized as a problem of
cascades for more than 15 years: it was found in hadronic cascades even
at low energy ($\approx$ 1 \GeVA) that the collision sequences and the
resulting distributions depend noticeably on the reference frame (LAB or
CMS) in which the simulation is run \cite{Kod84,Gyu97}.

The frame-dependence gets worse if one also makes the interaction
criterion dependent on the spatial distance of particles
\footnote{It may be worthwhile to point out an additional frame
          dependence inherent in any description involving partons: the
          \emph{parton concept itself} is frame-dependent, since it is
          well-defined in the infinite-momentum frame only.
         }.

If, on the other hand, we insist on \poin-covariance we run into trouble
because of the \emph{No Interaction Theorem} \cite{Cur63}, which asserts
that interacting particles cannot be described \poin-covariantly within
a classical 6N-dimensional Hamiltonian theory of point particles.

\subsection{\poin-Covariant Dynamics}
The solution of this problem chosen in the present work is the
\poin-Covariant Dynamics (PCD) approach described previously
\cite{Pet94}. In this approach the phase space is extended to $8N$
dimensions ($N$ is the number of particles). Position and momentum
vectors, ${r_{i}}^\mu(s)$ and ${p_{i}}^\mu(s)$ are parametrized by one
Lorentz-invariant parameter $s$ (which has no obvious physical meaning).

The particle motion and interactions are determined by the
Lorentz-scalar `Hamiltonian'
\begin{equation*}
  H = \sum_{i=1}^N \frac{{m_i}^2
    -{p_i}^2}{2m_i} \; +V  \;\;\; ,
\end{equation*}
where $V$ is a Lorentz-scalar pseudo-potential. The equations of motion
are
\begin{align*}
  \deriv{}{s}\;r_i(s) &= \{H,r_i\} = -\pderiv{H}{p_i} \eol
  \deriv{}{s}\;p_i(s) &= \{H,p_i\} = +\pderiv{H}{r_i} \;.\notag
\end{align*}

Since the evolution proceeds in the \emph{invariant} parameter   
$s$, every particle carries its own time $t_i = {r_i}^0(s)$. Note that
in this formalism particles are classically \emph{off-shell} when within
the range of the quasi-potential $V$.                            

\section{Dynamics of the model}
\label{sec:dyn}
Applying the concept of PCD to the \poin\ Covariant Parton Cascade
(PCPC), we simplify the dynamics by the following prescriptions: 
\begin{itemize}
\item Partons are free between collisions.
\item Binary interactions between partons $i$ and $j$ \emph{can} occur
  when their 4-distance $d_{ij}$ defined as
  \begin{equation}
    \label{eq:d-def}
    d_{ij} := \sqrt{-\left(\,r - \frac{r\cdot p}{p^2} p \right)^2}
              \;\;\; \text{with}\quad
    r := r_{i} -r_{j}  \;\;\; , \quad
    p := p_{i}+p_{j}
  \end{equation}
  has a minimum and \emph{will} occur if this distance is smaller than
  the interaction distance $\sqrt{\sigma_{\text{tot}}/\pi}$. Note that
  $d_{ij}$ is an Lorentz-invariant way of writing the spatial
  (3-)distance in the center of momentum frame.
\end{itemize}
Since the interaction criterion depends on Lorentz scalars only, the
ordering of interactions is \emph{frame-independent}.

\section{Initialization}

For setting up an initial state of the cascade, we define initial parton
distributions in two steps, successively resolving the nuclei into
nucleons and the nucleons into partons. First nucleonic Woods-Saxon
distributions (in momentum \emph{and} coordinate space) are used, and
the nucleons are boosted to the rapidities appropriate for the
particular collision. Then the flavor and momentum of each parton are
determined as follows: the parton momenta are

\begin{equation*}
  p^{\nu}
  = \left(\sqrt{(x{P_z}^{\text{nucl}})^{2}
      + {p_{\perp}}^{2}
      + \mu^{2} },
    \; p_{\perp}, \; x{P_z}^{\text{nucl}}\right)\;.
\end{equation*}
The longitudinal momenta $p_z = x{P_z}^{\text{nucl}}$ of the partons are
determined, together with their flavors, from the parton distribution
functions $f(x,Q^2)$ in the GRV94LO \cite{Gluck:1995uf} parametrization.
The distribution functions are evaluated at a fixed (and small)
${Q_0}^2$, i.e.\ at a fixed scale. The parton transverse momenta
$|p_\perp|$ are Gauss-distributed (mean value 0, width 0.3 GeV).

The one free quantity remaining is the parton's off-shell mass
$\mu =\sqrt{m^2-q^2}$. This is fixed by the fact
${|\vec{\beta}}^{\:\text{parton}}\,|\, =
\,\,\,|{\vec\beta}^{\:\text{\rule[-1.5ex]{0pt}{1em}nucleon}}\,|$\,.
This feature guarantees that partons originating from the same mother
nucleon move together initially, thus modeling the constraint that
initially partons are confined in the nucleons. Note how well PCD with
its concept of potential-dependent masses is suited to describe such
field-theoretic features in the classical framework of a cascade model.

The parton spatial coordinates $r_i$ are distributed exponentially in
the rest system of the nucleon. The boosted distributions are thus
Lorentz-contracted automatically, and ad-hoc modifications of the
spatial parton distributions (viz.\ ``Distributed Lorentz Contraction''
or ``boost invariant sea parton distributions'') that were proposed to
enhance the interaction rate are not needed in this model. Due to the
covariant formulation of our model they do not significantly influence
the cascade evolution.

\section{Parton Interactions}
\label{sec:int}

As stated in (Sec.~\ref{sec:dyn}), partons $i$ and $j$ may interact at
$s = s_{ij}$ if their invariant distance $d_{ij}(s)$ as defined in
(\ref{eq:d-def}) has a minimum value. At this moment the interaction
scale $Q^2$ (taken to be $-(t+u)$), and from that the relevant total
cross section $\sigma_{ij}$ for the two partons is determined. $Q^2$ is
not only the scale at which the running QCD coupling constant is
evaluated but also decides if the scattering is ``hard''. Then the total
cross section is determined from QCD in tree level approximation with
inclusion of massive quarks. Only if then
$d_{ij} \le \sqrt{\sigma_{ij}/\pi}$, the scattering actually occurs.

Although our model is based essentially on binary hard scatterings, some
non-perturba\-tive effects are included in the case in which the
interaction scale ${Q_h}^2$ of two partons is larger than the scale
${Q_0}^2$ at which the partons were initially resolved. In that case the
scale is adjusted via a ``DGLAP scale evolution''
\cite{Dokshitzer:1978hw,Dokshitzer:1977sg,Gribov:1972ri,%
      Lipatov:1975qm,Alt77,Ben86,Sjo85}.
This way $2\rightarrow n$ events are effectively incorporated. In the
picture below the scale evolution of parton $a$ from the initial scale
${Q_0}^2$ up to the interaction scale ${Q_h}^2$ is depicted. The
resolution scale is increased by successively radiated secondary
partons, and it is the final parton $b$ which is subjected to the binary
scattering process%
\footnote{Note the numbering: A larger scale means higher resolution,
          i.e.\ more partons.
         }.

\bigskip
\centerline{
            \unitlength.5cm \small \hspace*{1.5cm}
\begin{picture}(6.4,5.2)(0,0)
  \thicklines
  \put(5.74,4.54){\circle{1.2}}
  \put(5.74,4.54){\makebox(0,0){$d\sigma$}}
  \put(-1.5,2.5){\line(1,0){2}}
  \multiput(0,2.5)(0.4,0){8}{\line(1,0){0.15}}
  \put(2.5,2.5){\line(1,0){2}}
  \multiput(0.5,2.5)(2,0){3}{\line(1,-2){0.8}}
  \put(4.5,2.5){\line(1,2){0.8}}
  \put(-1.5,2.7){\makebox(1.5,0)[b]{{$a_n=a$}}}
  \put(2.6,2.7){\makebox(1.5,0)[b]{{$a_1=b_1$}}}
  \put(1.3,1.1){\makebox(0,0)[lb]{$c_n$}}
  \put(3.3,1.1){\makebox(0,0)[lb]{$c_1$}}
  \put(5.3,1.1){\makebox(0,0)[lb]{$c_0$}}
  \put(5,3.3){\makebox(0,0)[lt]{{$b_0=b$}}}
  \thinlines
  \put(-3.5,0.2){\vector(1,0){11.5}}
  \put(-0.75,0.1){\line(0,1){0.2}}
  \put(5.74,0.1){\line(0,1){0.2}}
  \put(8.1,0.2){\makebox(0,0)[l]{$Q^{{2}}$}}
  \put(-0.75,0.0){\makebox(0,0)[t]{$Q_0^{{2}}$}}
  \put(5.74,0.0){\makebox(0,0)[t]{$Q_h^{{2}}$}}
\end{picture}
           }

\section{Results}

As numerical results we show (Fig.\ 2.) rapidity and transverse momentum
distributions for central $\bar{p}\,$--$p$ and $Au$--$Au$ collisions at
RHIC energies (at $\sqrt{s}=200$\GeVA). They were calculated with a
Monte Carlo simulation based on the described model. Since the model in
its present version does not include a hadronization mechanism, we do
not compare our results with experimental data, but rather with a
different parton cascade model. We have chosen the above-mentioned VNI
code by Klaus Geiger \cite{Geiger:1998fq}, version 4.12 with
hadronization disabled (partons only), and soft interactions switched
off.

In the overall features VNI and PCPC agree reasonably well. For
$\bar{p}\,$--$p$ both models show a marked dip at midrapidity (as is to
be expected). For $Au$--$Au$ this dip is smeared out in our results, but,
surprisingly, in VNI it remains essentially unchanged. Note also the
significant differences in the transverse momentum distributions at low
$p_\perp$.

In comparing these results, it should be noted that in VNI the hadron
distributions obtained in runs with the hadronization mechanism switched
on, are markedly different from the parton distributions shown here.

\begin{figure}[tb]
  \begin{center}
    \unitlength1cm
    \small
    \begin{picture}(12,15.1)(0.1,-0.3)
      \put(-2, 7.5){
\begingroup%
  \makeatletter%
  \newcommand{\GNUPLOTspecial}{%
    \@sanitize\catcode`\%=14\relax\special}%
  \setlength{\unitlength}{0.1bp}%
\begin{picture}(3419,2051)(0,0)%
\special{psfile=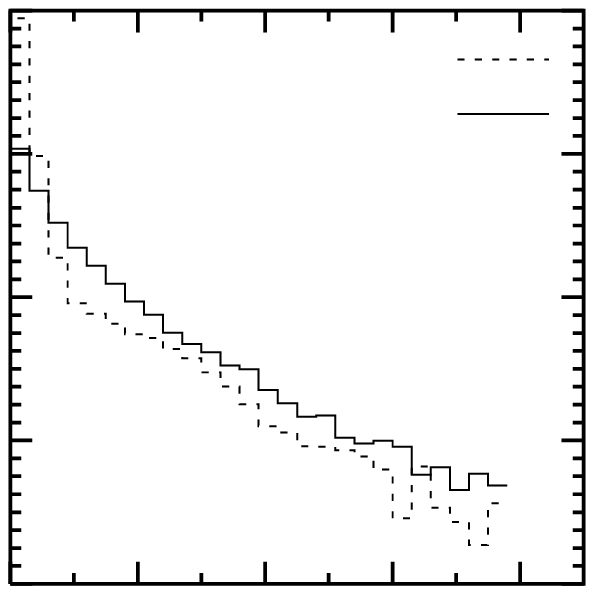 llx=0 lly=0 urx=684 ury=479 rwi=6840}
\put(1838,1653){\makebox(0,0)[r]{PCPC}}%
\put(1838,1810){\makebox(0,0)[r]{VNI}}%
\put(1508,1208){\makebox(0,0)[l]{$\textbf{\=pp}$}}%
\put(1425,50){\makebox(0,0){\normalsize$p_\perp$ (GeV)}}%
\put(300,1325){%
\special{ps: gsave currentpoint currentpoint translate
270 rotate neg exch neg exch translate}%
\makebox(0,0)[b]{\shortstack{$\displaystyle\frac{1}{Np_\perp} \frac{dN}{dp_\perp}$}}%
\special{ps: currentpoint grestore moveto}%
}%
\put(2068,200){\makebox(0,0){8}}%
\put(1701,200){\makebox(0,0){6}}%
\put(1334,200){\makebox(0,0){4}}%
\put(967,200){\makebox(0,0){2}}%
\put(600,200){\makebox(0,0){0}}%
\put(550,1951){\makebox(0,0)[r]{$10^{2}$}}%
\put(550,1538){\makebox(0,0)[r]{$10^{0}$}}%
\put(550,1126){\makebox(0,0)[r]{$10^{-2}$}}%
\put(550,713){\makebox(0,0)[r]{$10^{-4}$}}%
\put(550,300){\makebox(0,0)[r]{$10^{-6}$}}%
\end{picture}%
\endgroup
 }
      \put(6.8, 7.5){
\begingroup%
  \makeatletter%
  \newcommand{\GNUPLOTspecial}{%
    \@sanitize\catcode`\%=14\relax\special}%
  \setlength{\unitlength}{0.1bp}%
\begin{picture}(3419,2051)(0,0)%
\special{psfile=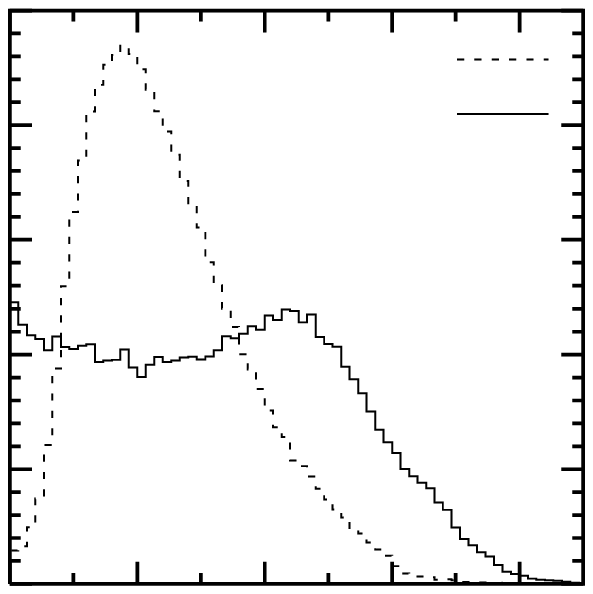 llx=0 lly=0 urx=684 ury=479 rwi=6840}
\put(1638,1653){\makebox(0,0)[r]{PCPC}}%
\put(1638,1810){\makebox(0,0)[r]{VNI}}%
\put(1308,1208){\makebox(0,0)[l]{$\textbf{\=pp}$}}%
\put(1225,50){\makebox(0,0){\normalsize$|\,y\,|$}}%
\put(100,1325){%
\special{ps: gsave currentpoint currentpoint translate
270 rotate neg exch neg exch translate}%
\makebox(0,0)[b]{\shortstack{$\displaystyle\frac{1}{N} \frac{dN}{d|y|}$}}%
\special{ps: currentpoint grestore moveto}%
}%
\put(1868,200){\makebox(0,0){8}}%
\put(1501,200){\makebox(0,0){6}}%
\put(1134,200){\makebox(0,0){4}}%
\put(767,200){\makebox(0,0){2}}%
\put(400,200){\makebox(0,0){0}}%
\put(350,1951){\makebox(0,0)[r]{0.05}}%
\put(350,1621){\makebox(0,0)[r]{0.04}}%
\put(350,1291){\makebox(0,0)[r]{0.03}}%
\put(350,960){\makebox(0,0)[r]{0.02}}%
\put(350,630){\makebox(0,0)[r]{0.01}}%
\put(350,300){\makebox(0,0)[r]{0}}%
\end{picture}%
\endgroup
 }
      \put(-2, -0.5){
\begingroup%
  \makeatletter%
  \newcommand{\GNUPLOTspecial}{%
    \@sanitize\catcode`\%=14\relax\special}%
  \setlength{\unitlength}{0.1bp}%
\begin{picture}(3419,2051)(0,0)%
\special{psfile=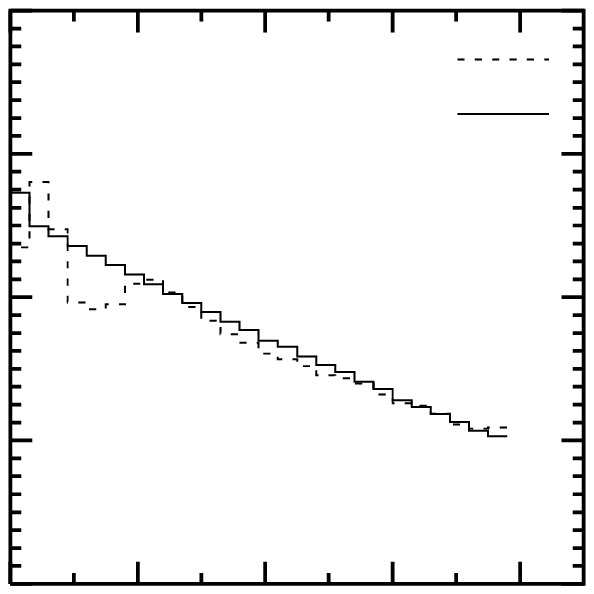 llx=0 lly=0 urx=684 ury=479 rwi=6840}
\put(1838,1653){\makebox(0,0)[r]{PCPC}}%
\put(1838,1810){\makebox(0,0)[r]{VNI}}%
\put(1508,1208){\makebox(0,0)[l]{\textbf{Au-Au}}}%
\put(1425,50){\makebox(0,0){\normalsize$p_\perp$ (GeV)}}%
\put(300,1325){%
\special{ps: gsave currentpoint currentpoint translate
270 rotate neg exch neg exch translate}%
\makebox(0,0)[b]{\shortstack{$\displaystyle\frac{1}{Np_\perp} \frac{dN}{dp_\perp}$}}%
\special{ps: currentpoint grestore moveto}%
}%
\put(2068,200){\makebox(0,0){8}}%
\put(1701,200){\makebox(0,0){6}}%
\put(1334,200){\makebox(0,0){4}}%
\put(967,200){\makebox(0,0){2}}%
\put(600,200){\makebox(0,0){0}}%
\put(550,1951){\makebox(0,0)[r]{$10^{2}$}}%
\put(550,1538){\makebox(0,0)[r]{$10^{0}$}}%
\put(550,1126){\makebox(0,0)[r]{$10^{-2}$}}%
\put(550,713){\makebox(0,0)[r]{$10^{-4}$}}%
\put(550,300){\makebox(0,0)[r]{$10^{-6}$}}%
\end{picture}%
\endgroup
 }
      \put(6.8, -0.5){
\begingroup%
  \makeatletter%
  \newcommand{\GNUPLOTspecial}{%
    \@sanitize\catcode`\%=14\relax\special}%
  \setlength{\unitlength}{0.1bp}%
\begin{picture}(3419,2051)(0,0)%
\special{psfile=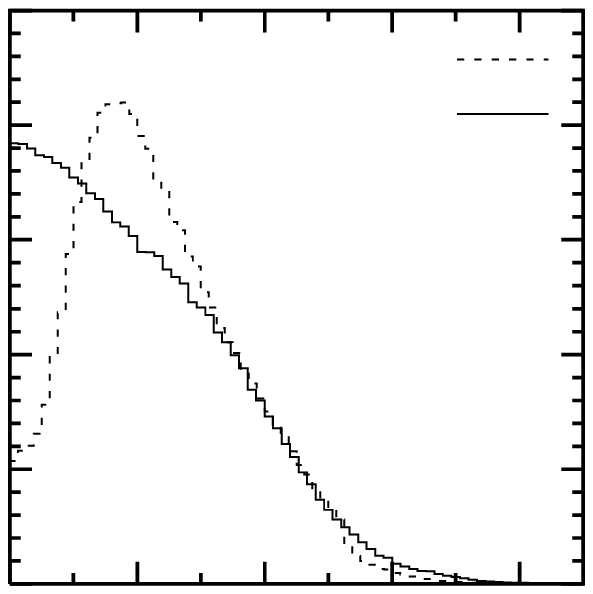 llx=0 lly=0 urx=684 ury=479 rwi=6840}
\put(1638,1653){\makebox(0,0)[r]{PCPC}}%
\put(1638,1810){\makebox(0,0)[r]{VNI}}%
\put(1308,1208){\makebox(0,0)[l]{\textbf{Au-Au}}}%
\put(1225,50){\makebox(0,0){\normalsize$|\,y\,|$}}%
\put(100,1325){%
\special{ps: gsave currentpoint currentpoint translate
270 rotate neg exch neg exch translate}%
\makebox(0,0)[b]{\shortstack{$\displaystyle\frac{1}{N} \frac{dN}{d|y|}$}}%
\special{ps: currentpoint grestore moveto}%
}%
\put(1868,200){\makebox(0,0){8}}%
\put(1501,200){\makebox(0,0){6}}%
\put(1134,200){\makebox(0,0){4}}%
\put(767,200){\makebox(0,0){2}}%
\put(400,200){\makebox(0,0){0}}%
\put(350,1951){\makebox(0,0)[r]{0.05}}%
\put(350,1621){\makebox(0,0)[r]{0.04}}%
\put(350,1291){\makebox(0,0)[r]{0.03}}%
\put(350,960){\makebox(0,0)[r]{0.02}}%
\put(350,630){\makebox(0,0)[r]{0.01}}%
\put(350,300){\makebox(0,0)[r]{0}}%
\end{picture}%
\endgroup
 }
    \end{picture}
    \caption{Comparison of the PCPC and VNI parton distributions for
             central collisions at $\sqrt{s}=200$\GeVA \,\,\,:
             $\bar{p}\,$--$p$ (top panels) and $Au$--$Au$ (bottom
             panels). The transverse momentum distributions
             $\frac{1}{Np_\perp}\, \frac{dN}{dp_\perp}$ are displayed on
             the left (cut: $|y|<1$), rapidity distributions
             $\frac{1}{N}\frac{dN}{d|y|}$ on the right.
            }
    \label{fig:rap}
  \end{center}
\end{figure}

\section{Summary}
The Poincar\'e-Covariant Parton Cascade Model (PCPC) as presented here
is still incomplete. In this stage, it incorporates the following
features:
\begin{itemize}
\item new cascade-type microscopic simulation model of
  Nucleus-Nucleus Collisions at RHIC energies
\item basic quantities are partons (quarks and gluons) moving in
  8N-dimensional phase space according to Poincar{\'e}-Covariant
  Dynamics (PCD)
\item Interactions: hard parton-parton scatterings with
  cross sections determined from perturbative QCD with a
  DGLAP scale evolution.
\end{itemize}

What is left to do next is:
\begin{itemize}
\item to include a hadronization mechanism
\item to add a thermodynamic analysis of results in space and time
\item to include shadowing effects, e.g.\ by using the EKS98
  parametrization of parton distribution functions \cite{Esk99}.
\end{itemize}
These are currently under development.

\bigskip\noindent
The PCPC code [in C$^{++}$] is obtainable from\\
\indent\texttt{http://hix.physik.uni-bremen.de/tkp}.


\end{document}